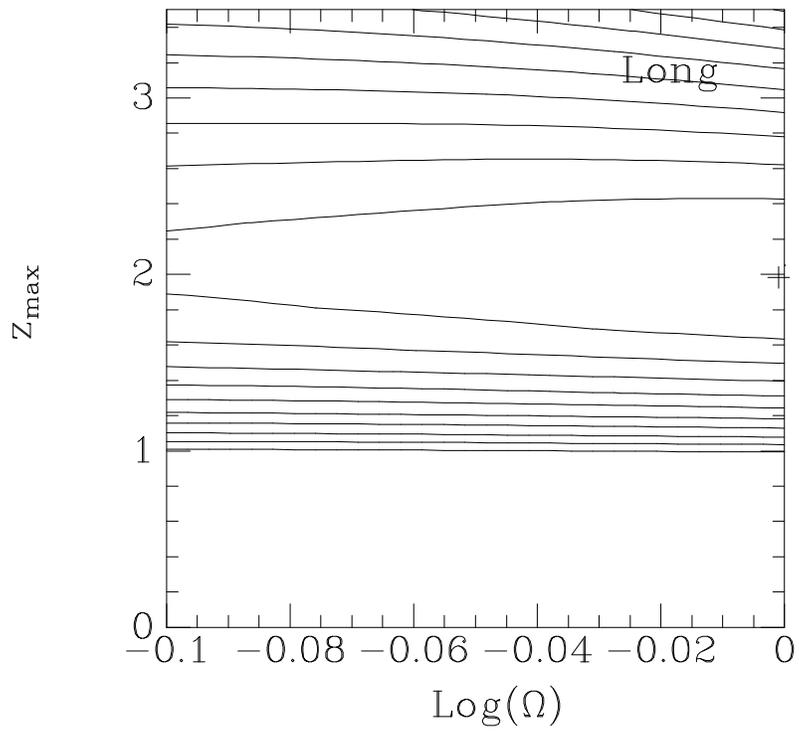

Fig. 1

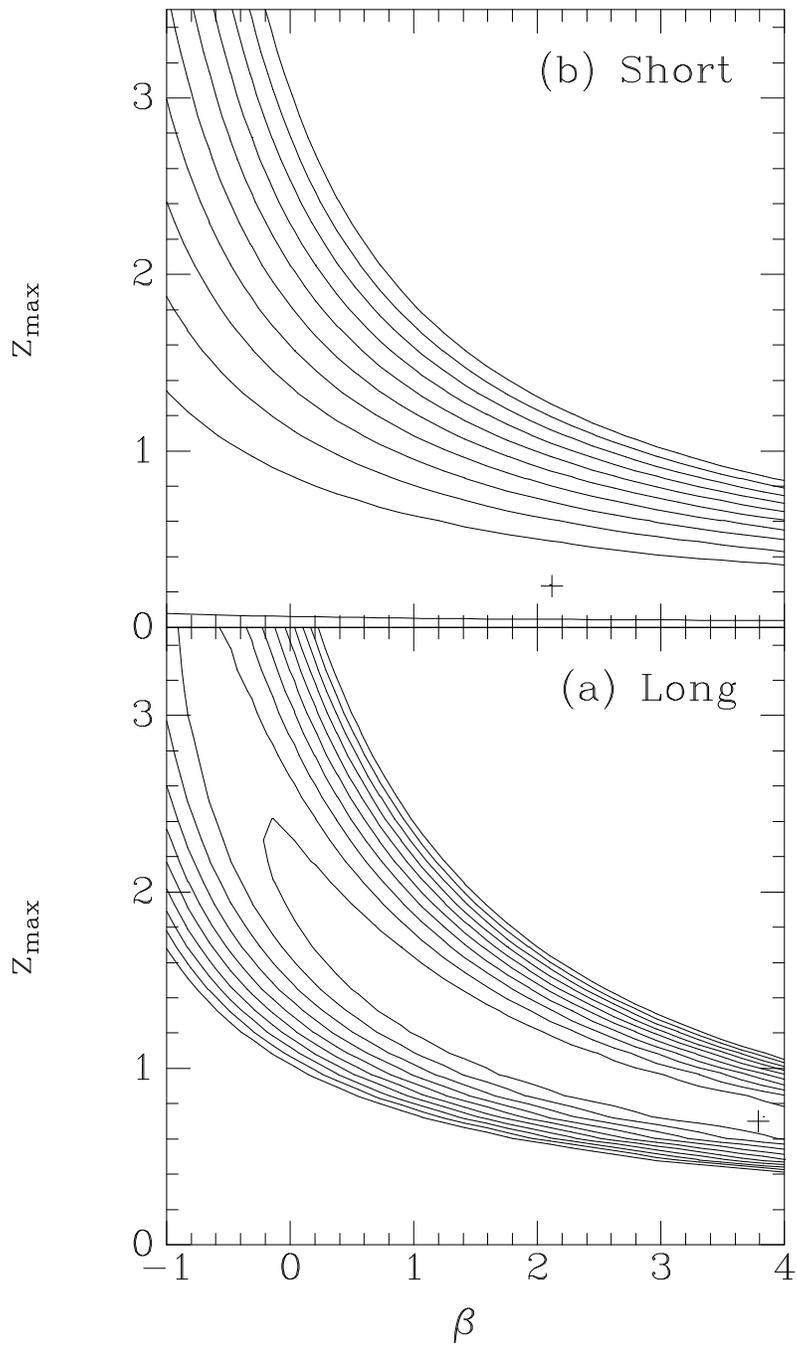

Fig. 2

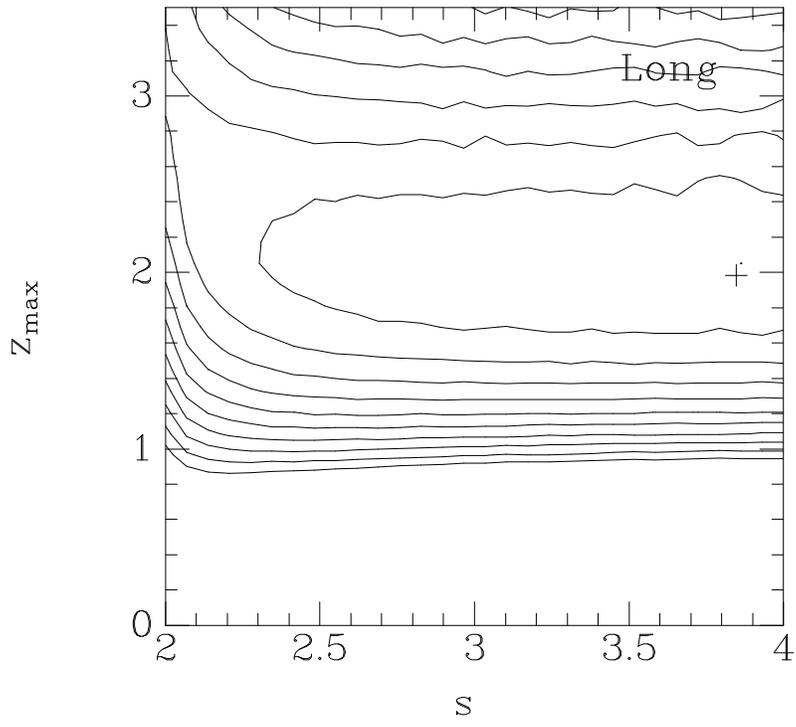

Fig. 3

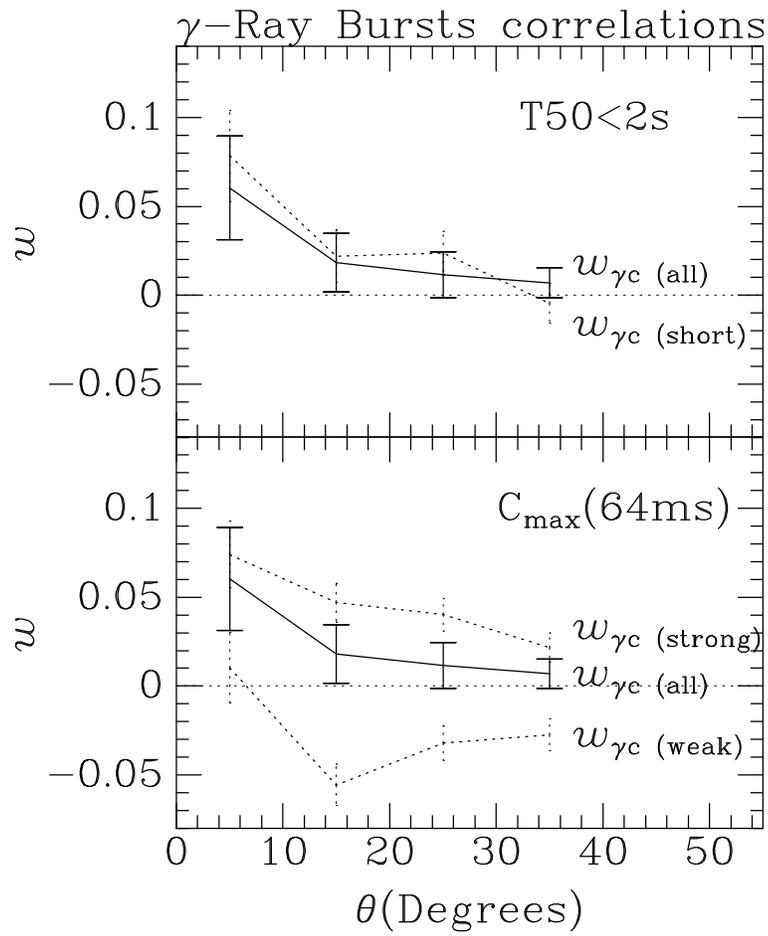

Fig. 4

# NEW EVIDENCE FOR THE COSMOLOGICAL ORIGIN OF $\gamma$-RAY BURSTS

Ehud Cohen, Tsafrir Kolatt and Tsvi Piran

*Racah Institute for Physics, The Hebrew University, Jerusalem, Israel 91904*

**Abstract**

We compare the burst distribution of the new (2B) BATSE catalogue to a cosmological distribution. We find that the distribution is insensitive to cosmological parameters such as $\Omega$ and $\Lambda$ and to the width of the bursts luminosity function. The maximal red shift of the long bursts is $\approx 2.1$ (assuming no evolution) while $z_{max}(long)$ of the short bursts is significantly lower $z_{max}(short) \approx 0.5$. In agreement with this relatively nearby origin of the short burst we find an indication that these bursts are correlated ($\geq 2\sigma$ level at 10°) with Abell clusters. This is the first known correlation of the bursts with any other astrophysical population and if confirmed by further observations it will provides additional evidence for the cosmological origin of those bursts.

*Subject headings:* gamma-rays: bursts — Cosmology

## 1  INTRODUCTION

The recent observation of the BATSE experiment on the COMPTON-GRO observatory have demonstrated, quite convincingly, that $\gamma$-ray bursts (GRBs) originate from cosmological sources (Meegan, 1992). The distribution of the count rate of the bursts in the BATSE 1B catalogue agrees well a cosmological distribution (Piran, 1992; Mao & Paczyński, 1992; Dermer, 1992; Schmidt, 1992; Wickramasinghe *et al.*, 1993, Mao, Narayan & Piran, 1994). Additionally there is a preliminary evidence for the predicted (Piran, 1992, Paczyński, 1992) correlations between the duration, the strength and the hardness of the bursts (Norris, *et al.*, 1994). With the release of the new (2B) BATSE catalogue (Meegan, *et al.*1994) we turn here to a new analysis of the data, which provides some new evidence for the cosmological origin.

Last year Kouveliotou *et al.*(1993) have shown that GRBs are bimodal in duration with short ($\leq 2$ s) and long ($> 2$ s) bursts. Lamb, Graziani & Smith (1993) and Mao *et al.*(1994) have shown that each sub population agrees with a cosmological distribution. Mao, Narayan

& Piran (1994) have shown that while BATSE's 1024ms channel is more sensitive to the long bursts and the 64ms channel is more sensitive to the short bursts, overall BATSE is more sensitive to the long burst which are detected to larger distances. At the same time both the long and the short bursts display a surprisingly similar instantaneous peak luminosity (to within a factor $\sim 2$).

We calculate the likelihood function and show that the observed count distribution in the 2B catalogue agrees with a cosmological distribution. The existence of the two sub-classes (with different sensitivities) is an important factor that we include in this analysis. It would be meaningless to combine both sub-classes to one distribution we treat each sub-class separately. Our results support the previous conclusion that the two sub-classes are observed to different distances. In the second part of the paper we calculate the cross-correlation between GRBs and Abell clusters. We find (at $\sim 2-3\sigma$ level) a cross-correlation for the whole GRBs distribution with Abell clusters and a similar cross-correlation between the sub-groups of strong GRBs and of short GRBs (both sub-groups are expected to be nearer to us).

## 2  Basic Model

We review here, briefly, the basic features of a cosmological source count distribution. Details can be found in (e.g. Weinberg, 1973) for a general population of sources and in (e.g. Piran, 1992) for GRBs. We consider a detector with with a fixed energy range, $\Delta D$ that operates for a fixed time span, $\Delta t$, and a source with an energy spectrum: $N(E) \propto E^{-\alpha}$. The count rate at the detector from a source at a redshift $z$ is:

$$C(\tilde{L}, z) = \frac{\tilde{L}(1+z)^{2-\alpha}}{4\pi d_l^2(z)} \qquad (1)$$

where $\tilde{L}$ depends on the luminosity of the source, $L$, in the relevant energy range, $\Delta E$, on the average energy $\bar{E}$ and on the observation time, $\Delta T$: $\tilde{L} = L(\Delta E)\Delta t/\bar{E}$. Apart from normalization which does not vary from one source to another $\tilde{L} \propto L$ and hence we will delete the ~ in the following. $d_l(z)$ is the luminosity distance and for a given $z$ it is a function of $\Omega$ and $\Lambda$.

Given a population of sources with a luminosity function $n(L)$ and with a rate of events $n(z)$ (that is at redshift $z$ we have $n(z)$ events per unit proper volume and unit proper time). The number, $N(>C)$, of observed events with an observer count rate larger than $C$ is:

$$N(>C) = \int_0^\infty n(L)dL \int_0^{z(C,L)} \frac{n(z)}{(1+z)}dz, \qquad (2)$$

where $z(C,L)$ is obtained by inverting Eq. ??. This theoretical $N(C)$ distribution depends on the cosmological parameters: $\Omega$ and $\Lambda$, on the source parameters: $\alpha$, $n(z)$ and $n(L)$. Following Schafer *et al.*(1992) we assume that all the sources have a spectral index of $-1.5$. This is a crude approximation, but lacking specific spectral indices for the individual bursts this is the best that can be done at present. $n(z)$ and $n(L)$ are functions which for simplicity we characterize in a few simple ways using one or two parameters. We will characterize the distributions in terms of $z_{max}$, the maximal $z$, from which the detector can detect the sources. For a standard candle distribution, $n(L) = \delta(L - L_0)$, $z_{max}$ is given by: $C_{min} = L_0(1+z_{max})^{-\alpha+2}/4\pi d_l^2(z_{max})$. Otherwise it is defined as an average over the distribution, for example $z_{max}(\langle L \rangle)$. In any case $z_{max}$ is a feature of the combination of the detector the source population and the real source distribution does not end at $z_{max}$.

## 3  Maximal Likelihood analysis

The 2B catalog available in the public domain contains a list of all gamma-ray bursts that triggered the BASTE detectors up to March 9th 1993. The catalog contains the ratio $C_{max}/C_{min}$ for 412 bursts, where $C_{max}$ is the maximum count rate and $C_{min}$ is the detection threshold and Duration table for 434 bursts (Meegan et al. 1994). For a variety of reasons, BASTE has a variable background as a function of time, so that the detection threshold $C_{min}$ does not remain constant. In order to have a more uniform sample we select a constant threshold $C_{cut}$ ($C_{cut}(1024ms) = 287$ and $C_{cut}(64ms) = 72$), and prune the data to contain only bursts which satisfy $C_{min} \leq C_{cut}$ and $C_{max} \geq C_{cut}$ (Mao, Narayan & Piran 1993). Following Kouveliotou et al. (1993) we define "short bursts" as having $\delta t_{90} < 2s$ and "long bursts" as having $\delta t_{90} > 2s$. In order not to exclude bursts which don't have a duration record we have used the relation, based on the behavior of 64msec counting and 1024msec counting (Mao, Narayan & Piran 1993), for long bursts $(C_{max}/C_{cut})_{1024}/(C_{max}/C_{cut})_{64} > 1$ and the opposite relation for the short bursts. Using this relation we add 15 more bursts to our set. After excluding bursts which don't satisfy with all the previous requirements we have 238 long bursts, and 50 short bursts. We have calculated the likelihood function for the multidimensional phase space discussed before. In the following discussion we describe several two dimensional cuts through this phase space.

### 3.1  Insensitivity to the cosmological parameters: $\Omega$ and $\Lambda$

Figure 1 depicts the contourlines of likelihood function for the long bursts distribution in the ($\Omega$,$z_{max}$) plane. The likelihood function is insensitive to $\Omega$ in the range (0.1-1). This follows from the relatively modest $z_{max}$ value of the currently observed bursts. At these values the main effect of $\Omega$ on Eq. ?? is to change the length scale. It hardly changes the overall shape of the $N(C)$ distribution. Similarly we find (figure not shown) that the likelihood function is insensitive to a cosmological constant $\Lambda$ (in the range 0-0.9). Following these results we present in the rest of the paper representative likelihood functions for cosmologies with $\Omega = 1$ and $\Lambda = 0$.

### 3.2 Source Evolution

The cosmological evolution could be masked by an intrinsic evolution of the sources. To examine the effect of intrinsic evolution of the number count we have introduced a $z$ dependent local event rate: $n(z) = n_0(1+z)^{-\beta}$. We describe, in Figure 2, the likelihood function in the $(z_{max}, \beta)$ plane. A distribution with no intrinsic $z$ dependence is equivalent to a distribution with an intrinsic $z$ dependence with a lower $z_{max}$. This result confirms the earlier results of Piran (1992) which were based on the $\langle V/V_{max} \rangle$ of the 1B catalogue (Note that Piran (1992) describes the intrinsic evolution is described as $n(t) = n_0 t^{\beta'}$ and for low $z_{max}$ we have $\beta' \approx -3\beta/2$).

### 3.3 Luminosity Function

To examine the possible effects of a luminosity function consider a source distribution with a luminosity function of the form $n(L) = n_0(L/L_0)^{-s}$ for $L > L_0$ (and $N(L) = 0$ otherwise). Figure 3 depicts the likelihood function in the $(s, z_{max}(\langle L \rangle))$ plane. The likelihood function is quite insensitive to $s$ as long as $s \gtrsim 2.2$ (note that $\langle L \rangle$ diverges for $s \to 2$). This shows that, in contrast to previous claims (Mao & Paczyński, 1992), a narrow luminosity function is not needed in order to obtain the observed GRB distribution and GRBs are not necessarily standard candles. The insensitivity to the width of the luminosity function was confirmed when we tried several other luminosity functions, including a combination of two standard candles, a broken power law and a rising power law with an upper cutoff.

### 3.4 Long vs short sub-populations

Figure 2 also depicts the likelihood function for the short bursts sub-population. In agreement with Mao, Narayan & Piran (1994) we find that the short burst sub-class is nearer than the long burst sub-population. With no source evolution we find that $z_{max}(short) <$ 1.6 at a 1% confidence level with a maximal likelihood around $z_{max}(short) = 0.4$. The likelihood goes down to $\approx 20\%$ of the maximum for $z \approx 0$, hence a nearby (non-cosmological) population cannot be ruled out. This is probably due to the relatively low number of bursts (50) in this sub-group. For the long bursts we obtain: $1.4 < z_{max}(long) < 3.1$ at a 1% confidence level and a maximal likelihood around $z_{max}(long) \approx 2.1$. If we recall that the 1024ms channel, that is detecting the long bursts is more sensitive than the 64ms channel that is detecting the short bursts we find (again in agreement with Mao, Narayan & Piran 1994) that both populations have a comparable maximal luminosity.

## 4  Cross correlation with Rich Clusters

We did not find any significant auto-correlation of GRBs in the 2B catalogue. This agrees with previous results (Hartman and Blumenthal, 1989) Since GRBs are distributed over cosmological distances ($z \approx 1$) one could not expect a cross correlation with galaxies which are relatively nearby. Indeed we did not find any correlation with IRAS galaxies or with galaxies in any other optical catalogue. However, Abell clusters (Abell 1958, Abell, Corwin & Olwin 1989) are observed up to $z \approx .15 - .2$ and this leads to a possible cross - correlation, in particular with strong bursts which presumably originate nearby. We present here preliminary results about the cross correlation between the GRBs and rich Abell clusters. Detailed analysis with respect to various classes of Abell clusters is described elsewhere (Kolatt et al., in preparation).

We carried out the cross correlation with the rich clusters data in the traditional way. Let $N_{gr-cl}(\theta)$ be the number of pairs of GRBs and clusters separated by the angle $\theta$ where for the $i'th$ bin $\theta_{i-1} < \theta < \theta_i$ and let $N_{gr-po}(\theta)$ be the number of pairs of GRBs and Poissonian distributed particles selected according to the same criteria as for the clusters. Then the angular cross-correlation function is defined as

$$1 + w(\theta) = \frac{N_{gr-cl}(\theta)}{N_{gr-po}(\theta)} \frac{n_{po}}{n_{cl}}, \qquad (3)$$

whereas $n_{cl}$ and $n_{po}$ are the number densities of the clusters and Poisson catalogs respectively. In order to account for the different selection criteria we selected GRBs sources with the probability $1 - P(\delta)$, $\delta$ being the terrestrial declination, where

$$P(\delta) = \frac{\#\ of\ observing\ days\ (\delta)}{Maximal\ \#\ of\ observing\ days} \qquad (4)$$

This procedure prevents us from compensating for the selection $\phi(\delta)$ by multiplying by $\phi^{-1}(\delta)$ and from increasing the noise level in poorly sampled declinations. The selection criteria for the clusters were considered by applying the same selection to the Poisson catalog. We followed Scaramella et al.(1991) and used

$$\phi_{cl}(|b|, \delta) = \text{dex}\left[-A(cosec(|b|) - 1)\right] \times \text{dex}\left[-B(sec(|\zeta|) - 1)\right], \qquad (5)$$

$\zeta$ being the zenithal angle, $\zeta = \delta - \delta_{telescope}$, with the appropriate parameters as quoted there for the north and south samples. On top of these selection criteria we restricted ourselves to Galactic latitude $|b| > 30°$ for both the GRBs and the clusters.

A main advantage of the cross correlation method is that it allows to reduce the noise level appreciably. In the limit where $n_{po} \gg n_{cl}$, the error in the $w(\theta)$ due to the Poisson error is:

$$\frac{\Delta w}{1+w} = \frac{1}{\sqrt{N_{gr-cl}}}, \qquad (6)$$

for each $\theta$ bin. The Poissonian errors should still be amplified due to cluster - cluster correlations and possible GRBs auto correlation. Figure 4. shows the cross correlation between 3616 rich clusters and the GRBs (solid line), the bins' size was determined by the GRO angular resolution and was set to $10°$. The errors shown for the solid line are *not* given by Eq. 6. Instead, we try here to reject the hypothesis of the GRBs being Poissonianly distributed and that the entire cross correlation signal is contributed by the clusters auto correlation. We created an ensemble of Poissonian distributed GRBs (same selection and number), from which we got the average pair number in each bin as expected for a Poissonian model. The error then, was taken to be $\Delta w = \sqrt{1+u}/\sqrt{<Npairs>}$, $u$ being an upper limit to the two dimentional $J_3$ integral ($u \simeq 4$). The amplifying factor takes care of the fact that the true number of independent pairs is closer to the GRBs - super clusters pairs rather than the GRBs - clusters pairs. A more elaborated error analysis is carried out elsewhere (Kolatt *et al.*, in preparation). No ensemble average cross correlation between the Poissonian model and the clusters was detected. One notices that we can reject the hypothesis of GRBs being randomly distributed on the $\geq 2\sigma$ level at the $10°$ bin.

We proceed by eliminating the presumably non overlapping GRBs and cross correlate only the remaining sources. This procedure enables a higher correlation signal level by preventing the contamination due to far sources. However, at the same time the noise level increases too, due to the reduction in pair numbers. The elimination can be done in two ways. We divide the GRBs to strong and weak bursts and compare the cross correlations of the two groups with rich Abell clusters. The sum of the two groups *does not* result to the entire sample number since $C_{max}$ is not registered for all GRBs. We also divide the GRBs to long and short bursts, assuming short bursts are significantly nearer than longer bursts. For comparison, we calculate again the cross-correlation function, this time with the previously eliminated sources to check consistency of the result with null correlation. Figure 4 shows the cross correlation of the strong sources (defined by $C_{max} > C_{max}(median)$ for the $64 ms$ Channel) with the rich clusters (lower panel). The errors shown, here, are the Poissonian errors (Eq. ??) and should be used only as an indication. The actual errors derived from a Monte Carlo simulation are larger. The orthogonal set of weak sources is shown too. We would like to emphasize that there might be no statistical significant to the signal obtained for each bin, but the bifurcation of the two subsets is absolutely consistent with the two distinct populations assumption. Figure 4 (upper panel) also shows the same test for sources with short duration ($< 2s$), again with Poissonian errors (Eq. **??**). This test turns out again to be consistent with the picture described earlier about two clearly distinguished populations for the GRBs and with the fact that the short bursts are typically nearer than long ones.

## 5  Conclusion

Our results demonstrate that the GRB population is cosmological. Unlike preliminary hopes, (Trimble, 1994) we find that the burst distribution is not suitable to serve as a cosmological tool that will distinguish between cosmological models with different cosmological parameters. It is possible, but quite unlikely, that much more sensitive detectors that will observe the bursts to higher $z$ could do that. This is doubtful since cosmological evolution could easily mask the effects of the background geometry. We also find that the distribution is rather insensitive to the width of the luminosity function and several luminosity functions with different width agree with the data.

We also confirm the earlier conclusion (Kouveliotou *et al.*. 1993) the GRBs distribution is divided to long and short bursts. BATSE is more sensitive to the long bursts. If there is no burst evolution and the bursts are standard candles then the maximal red shift up to which the long bursts are observed is $z_{max}(long) = 2.1^{+1.1}_{-0.7}$ this corresponds to an event rate of $2.3^{-0.7}_{+1.1} \cdot 10^{-6}$ events per galaxy per year where we estimated the detection efficiency of BATSE as 0.3 and we used a galaxy density of $10^{-2} h^3$ Mpc$^{-3}$ (Kirshner *et. al*, 1983). The rate per galaxy is independent of $H_0$ and is only weakly dependent on $\Omega$. The typical energy of a burst is $7^{+11}_{-4} \cdot 10^{50} F_{-7}$ergs (where $F_{-7}$ is the total observed fluence in $10^{-7} ergs/cm^2$) for $\Omega = 1$. Those numbers vary slightly if the bursts have a wide luminosity function. The rate increases and the luminosity decreases if there is a positive evolution of the rate of bursts with time. The short bursts are observed only up to a much nearer distances: $z_{max}(short) = 0.4^{+1.1}$ (again for standard candles and no source evolution) which corresponds to a comparable rate of $6.3^{-5.6} \cdot 10^{-6}$ events per year per galaxy and typical energy of $3^{+39} \cdot 10^{49} F_{-7}$ergs. For the short sub-class, which have only 50 members, there is no significant lower limit on $z_{max}$.

Recently Norris *et al.*(1994) found an increase in the duration of weak bursts by a factor of 2 compared to short bursts. They interpret this as evidence to cosmological red shift and conclude that the weak bursts originate at $z \approx 1$. Their analysis applies to long bursts (longer than 1.5 sec) which, according to our model originate (with standard candles and no source evolution) from significantly further distances. This result can be reconciled with our analysis if there is a significant source evolution (see Fig. 2) or if there is a sufficiently wide luminosity function (the maximal likelihood of the latter case is however only 3% of the maximal likelihood of $z \approx 2.1$).

We also found an indication for positive cross correlation of GRBs with rich Abell clusters. Abell clusters extend up to $z \approx .15 - .2$ and correspondingly this cross-correlation is higher for the stronger half of the bursts (those being nearer) and with the short bursts (again those being a nearer sub-population). The cross-correlation is the first association of GRBs with any other population of astronomical objects. It supports, of course, the hypothesis of cosmological origin of the bursts. It does not mean, however, that the GRBs originate necessarily in rich clusters. As it is consistent with the picture in which the bursts originate in galaxies which, in turn, are strongly correlated with Abell clusters. This cross-correlation indicates that there is some overlap between GRBs and Abell clusters. This is self evident for the short bursts for which we found $z_{max}(short) \approx 0.4$. If the long bursts contribute significantly to the cross-correlation this suggests that long GRBs are not observed up to $z_{max} = 2$ but to a lower $z_{max}$ value. Just like the results of Norris *et al.*(1994) this can be reconciled with the $N(C)$ distribution if there is a source evolution which reduces $z_{max}$.

This research was supported by a BRF grant to the Hebrew University.

# FIGURE CAPTIONS

**Figure 1.** : Contourlines of likelihood function in the $(\Omega, z_{max})$ plane for standard candles sources and with no source evolution. The contours are at likelihood levels 33%, 10%, 3.3% 1% etc.. of the maximum. The + sign marks the place where the likelihood has a maximum. The likelihood is insensitive to $\Omega$.

**Figure 2.** : Contourlines of likelihood function in the $(\beta, z_{max})$ plane for standard candles sources and $\Omega = 1$. (Bottom: Long bursts; Top: short bursts).

**Figure 3.** : Contourlines of likelihood function in the $(s, z_{max}(\langle L \rangle))$ plane for $\Omega = 1$ and no intrinsic evolution.

**Figure 4.** : Cross correlation of GRBs and Abell clusters. Solid line (in both panels) is for the entire GRB population, with realistic upper limit errors for a Poissonian *model* of GRBs (see text). The upper panel depicts cross correlation of GRBs for which $T50 < 2\,ms$ (dotted line). The Lower panel shows the split of the cross correlation when subsets according to strength are used. The errors in these two cases are for simple Poissonian distribution (Eq. 6) and serve as a lower limit.